\documentclass[english,pdftex]{MetroArchaeo2019}
\usepackage{hyperref}
\usepackage{graphicx}
\usepackage{amssymb}
\usepackage[font={normalsize,it}]{caption}
\usepackage{xcolor}
\usepackage{amsmath}
\usepackage[symbol*]{footmisc}

\title{Measurement and analysis of visitors' trajectories in crowded museums$^*$}
\author[1]{Pietro Centorrino}
\author[2]{Alessandro Corbetta}
\author[3]{Emiliano Cristiani}
\author[4]{Elia Onofri}
\affil[1]{
    Dipartimento di Fisica, Sapienza Universit\`a di Roma, Rome, Italy
}
\affil[2]{
    Department of Applied Physics, Eindhoven University of Technology, Eindhoven, The Nethederlans
}
\affil[1,3,4]{
    Istituto per le Applicazioni del Calcolo, Consiglio Nazionale delle Ricerche, Rome, Italy, e.cristiani@iac.cnr.it
}

\begin{document}
\maketitle
\thispagestyle{imekopage}
\pagestyle{empty}
\begin{abstract}
We tackle the issue of measuring and analyzing the visitors' dynamics in crowded museums. We propose an IoT-based system -- supported by artificial intelligence models -- to reconstruct the visitors' trajectories throughout the museum spaces. Thanks to this tool, we are able to gather wide ensembles of visitors' trajectories, allowing useful insights for the facility management and the preservation of the art pieces.
Our contribution comes with one successful use case: the Galleria Borghese in Rome, Italy.
\footnote[0]{
    $^*$Results presented in this paper are achieved under the project \textit{Management of flow of visitors inside the Galleria Borghese in Rome}, supported by Ministry of Cultural Heritage and Activities and Tourism, Galleria Borghese, and Istituto per le Applicazioni del Calcolo of National Research Council of Italy.
    Project's Principal Investigators are Marina Minozzi (Galleria Borghese) and Roberto Natalini (IAC-CNR).

    E. Cristiani also acknowledges the Italian Minister of Instruction, University and Research to support this research with funds coming from PRIN Project 2017 No.\ 2017KKJP4X entitled \textit{Innovative numerical methods for evolutionary partial differential equations and applications}.

    A. Corbetta also acknowledges the support of the Talent Scheme (Veni) research programme, through project number 16771, which is financed by the Netherlands Organization for Scientific Research (NWO).}
\end{abstract}


\section{Introduction}
Visitors' flow management is a central issue for world-leading museums. Museum managers are constantly challenged by the need of maximizing the number of visitors while ensuring individual  safety and comfort, and preserving inestimable collections~\cite{Elmamooz1, Kontarinis1}. 
Measuring and understanding the visitors' behavior is crucial to tackle this challenge. Yet, this involves highly complex issues: continuous and reliable data acquisition, complexity reduction, modelling physical and psychological aspects of crowd motion. 

In this paper we present an innovative, IoT-based system for measuring and understanding the visitors' dynamics. The system is inexpensive and non-invasive, as it is based on small beacons given individually to visitors at the entrance. The system enables accurate room-scale tracking of the visitors, close to real-time.

\section{CASE STUDY: GALLERIA BORGHESE}
The world-renowned Galleria Borghese in Rome, Italy, provided the environment for the development and test of our system. The museum is divided in two floors connected by a staircase and has 21 rooms. While being relatively small in size, the museum receives more than 500,000 visitors per year. To deal with the visitor flows and cope with the many historical, artistic and architectural constraints, the museum established a crowd management strategy based on scheduling entrance and exit times. Currently, the visits are organized as follows: tickets must be booked in advance and give access to the museum for a two-hour time slot. A small percentage of tickets, called ``last minute'', are instead sold 30 minutes after the beginning of a slot. At the end of each time slot, people are invited to leave, and the museum empties completely. 
Visits are not organized, in the sense that no obligatory exhibition path or predetermined sequence of rooms exist. This is done on purpose and it is useful to satisfy the different capacity limits of the floors.
The circular structure of the building, moreover, makes visits nonlinear: visitors return multiple times to the same room, either to reach the stairs or to admire artworks missed during the first visit of a room.

The data presented in this paper come from a measurement campaign lasted multiple days in the period Jul-Sept 2019,  which allowed the collection of about 900 visitors' trajectories, labelled manually in three sets of visiting experience: trajectories of visitors following a human guide (\emph{Guide} trajectories), trajectories of visitors supported by an audio guide (\emph{Audioguide} trajectories) and all the others (\emph{Normal} trajectories).

\section{IoT visitors tracking system}

We track anonymously visitors by means of Bluetooth Low Energy (BLE) radio beacons that are individually provided at the entrance and returned at the end of the visit  (for a similar approach, see, \emph{e.g.}~\cite{Ratti1}). Each beacon periodically broadcasts its identity. 
In every room, we deployed one or more receiving units, with minimum computing and data transmission capabilities (built with Raspberry Pi 3B+, in short RPi).
Due to technical hindrances, we were forced to consider a simplified planimetry that divided the first floor into 8 rooms (covered by 10 RPis) while aggregating the second floor in a unique room (covered by 4 RPis). Therefore we identify 10 rooms (by also considering the chance of being outside the museum) with 14 receivers.
The RPis capture the beacon identities and re-transmit them through a local network to a central server, along with the Received Signal Strength indicator (RSSi) and the timestamp of the event. 
By processing the data transmitted by all receiving units, the server estimates individual trajectories. In the next subsections we detail this reconstruction process.

\subsection{Trajectories Reconstruction}
The central server rebuilds the sequence of rooms of each visitors from the RSSi data. This data are generally highly noisy due, \emph{e.g.}, to electromagnetic disturbances, uneven sampling rate,  possible ambiguities in the readings from different rooms and limited network speed. 
As a first processing, we operate a preliminary data de-noising, by  
resampling over $m$ bins of time lenght $\Delta t = 10$ seconds.

If $n$ is the number of RPi stations, we therefore obtain a $n \times m$ matrix $R$ where each row represents the data gathered by an RPi and each column represents a time bin.

We now present three different approaches that we have developed in order to estimate the trajectories from $R$.

\subsubsection*{ArgMax-based trajectory reconstruction}

We evaluate, for each beacon and each time bin, the room in which the RSSi is the strongest.
The maximum of the arguments, from which ArgMax (AM) name comes from, thus consists in evaluating the maximum by columns of $R$.

\begin{figure}[t]
\centering
\includegraphics[width=0.95\linewidth]{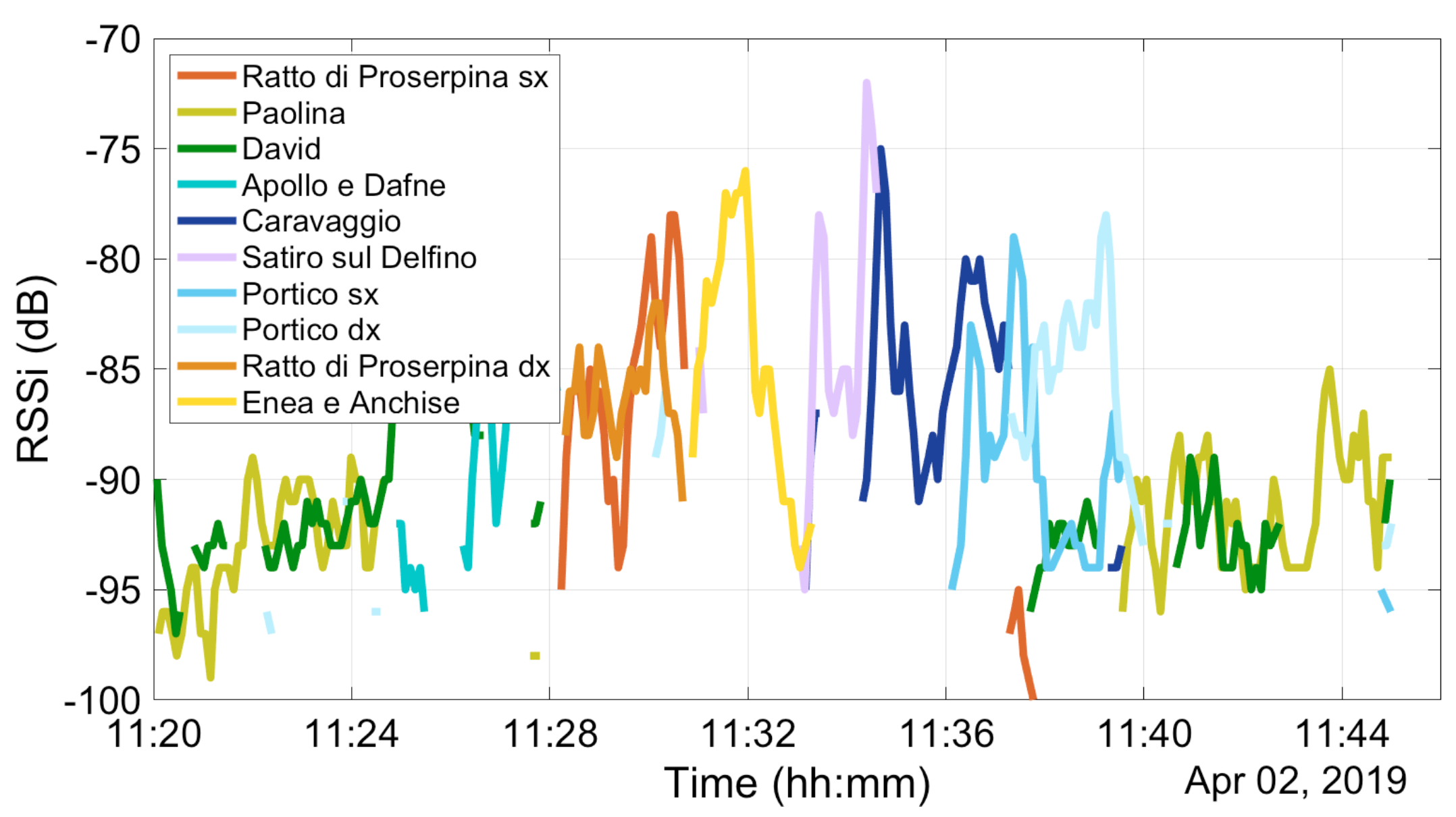}
\caption{Signal strength (RSSi) of a single beacon (visitor) after the application of both Moving Average and Normalization.}
\label{fig:ArgMax}
\end{figure}

\begin{figure}[t]
    \centering
    \includegraphics[scale=0.16]{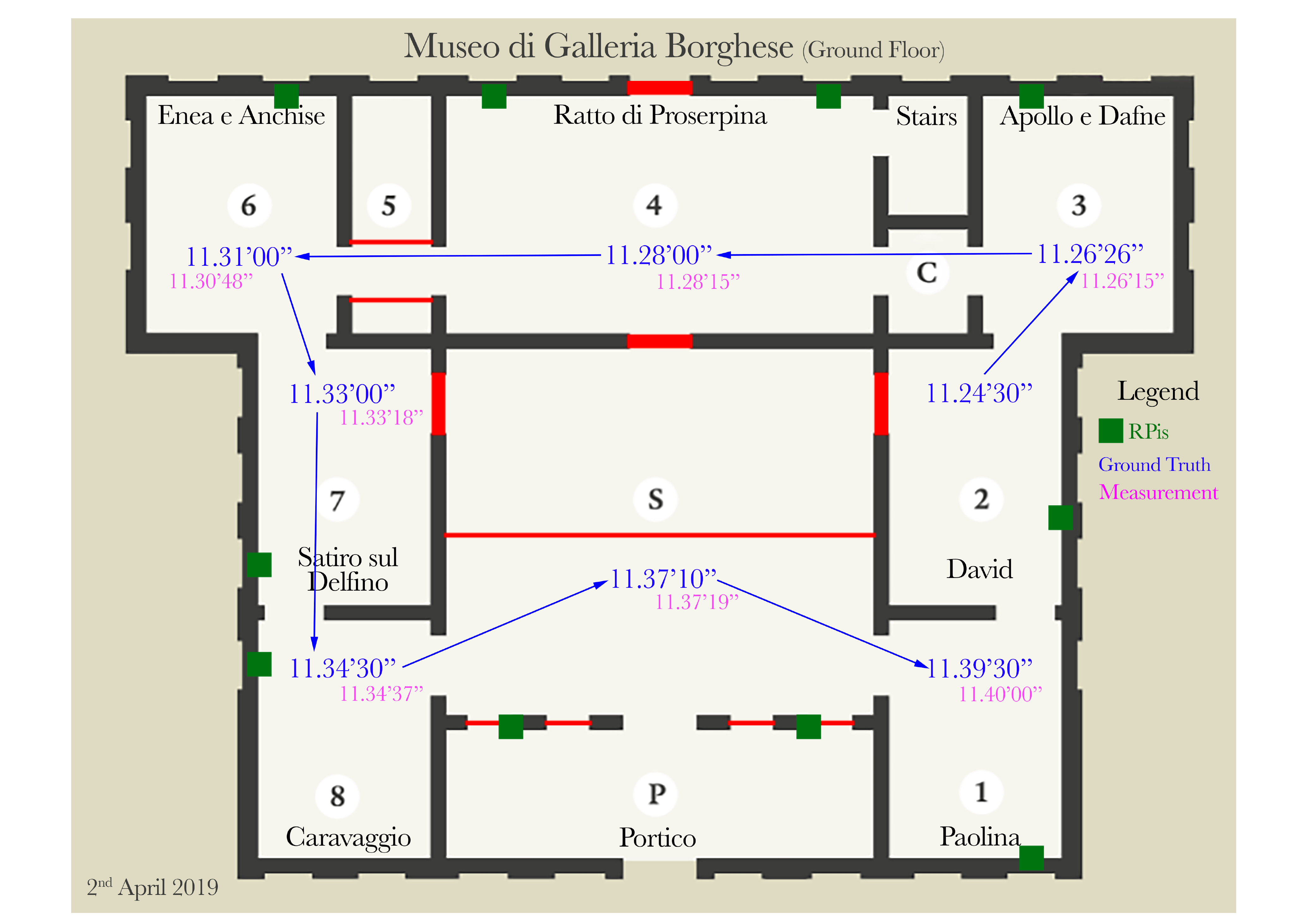}
    \caption{
        Trajectory of the visitor in Fig.\ref{fig:ArgMax} across the first floor of the museum. We report the measured entrance time for each room, compared to the corresponding ground truth recorded by the 
        researcher. 
    }
    \label{fig:DvsGT}
\end{figure}

This approach lays the basis for the comparisons we are going to make with two additional methods we now introduce. First, we enhance the AM method via two preliminary data refinement procedures that raise the accuracy of the result, as follows:
\renewcommand{\theenumi}{\roman{enumi}}
\begin{enumerate}
\item \emph{Moving average}: we select an ``history'' range $(\delta_-, \delta_+)$ in order to smooth the data from the measurement errors, and we apply a moving average to obtain $\tilde R$, defined as follows:
    \begin{equation}
        \tilde R_{i, \cdot} = \frac{\sum_{\ell=i-\delta_-}^{i+\delta_+} R_{\ell, \cdot}}{1 + \delta_+ + \delta_-}, \quad \delta_- < i \leq n - \delta_+.
    \end{equation}
    This enhances the strong relation between time and location by lowering data fluctuations. 
    A two-minute window ($(\delta_-, \delta_+) = (6, 6)$) proved to be the best compromise maintaining a satisfactory accuracy.
    
 \item \emph{Normalization}: we normalize the data by row (\emph{i.e.}\ by room in order to reduce the high variance of RSSi values and strongly penalize the rooms where the signal was not detected, building $\bar R$:
    \begin{equation}
        \bar R_{i, j} = \frac{\tilde R_{i, j} - \mu_i}{\sigma_i}
    \end{equation}
    where $\mu_i
    $ is the average by row and $\sigma_i
    $ is the standard deviation.
\end{enumerate}
The maximum by columns of $\bar R$ gives us better results, see Figures \ref{fig:ArgMax}-\ref{fig:DvsGT}. We will refer to this approach as Mobile Average (MA) method.


\subsubsection*{Machine Learning-based trajectory reconstruction}

Neural Networks (NNs) and Deep NNs have been widely used in many modern problems \cite{ML-NN-book}.
One of their best efforts is the ability to ``learn'' to solve a task like a human being would.
However, the problem of massive labelled data gathering needed for a correct effort is the main limit of this approach.

In particular a NN is built over a few parameters: a number of layers $L$ and a number of units $s^{(l)}$ per layer $l$. In our case the choice of these numbers was driven by the small number of labeled data we gathered ($L = 3$) and by the room-scaled resolution we choose to achieve ($s^{(3)} = 10$).

In order to limit the data needed for the network training, we combined different approaches:
\begin{enumerate}
    \item \emph{Threshold System}: 
    we firstly filter and clean the noisy data gathered from the beacons not yet entered inside the museum (randomly detected by any of the RPi station).
    \item \emph{Neural Network}: the main procedure performs the analysis of the position one bin at a time. By tracing the moving average approach described above, we evaluate the data of the current bin by considering the preceding $\delta_- = 6$ bins and the successive $\delta_+ = 6$ bins: This forces the choice of $s^{(1)}$ to $n \cdot (1 + \delta_- + \delta_+) = 14 \cdot 13 = 182$. This way the NN is able to learn approximately the most common transitions and the standard undesired fluctuations. The output of the NN is therefore a $10$-element vector $p^{(t)}$ per bin $t$ where the $i$-th value is the probability of a beacon to be at location $i$.
    \item \emph{Adiacency filter}: in order to determine the position of a beacon, a further step is made over $p$ by considering where the beacon comes from and where it moves to. The structure of the museum itself (\emph{i.e.}\ the connections between rooms) is here used to penalize the probabilities related to unfeasible transitions between locations that are not phisically connected.
\end{enumerate}

\subsection{Comparing the Approaches}
To compare methods (AM, MA and NN), we employ the accuracy measure, \emph{i.e.}\ the ratio between the correct predictions and the total number of samples analyzed.


For what concerns AM, we have evaluated the accuracy over a sample of $1500$ labelled bins, obtaining a result of $0.547$, enhanced to a value of $0.734$ with the MA method.

On the other hand, we have trained the NN with a $1e-4$-batch gradient descent over $5500$ labelled bins ($20000$ iterations from random generated weights) and we have tested its efficiency over a disjoint Test Set made of $1000$ data, achieving an accuracy of $0.858$.

\begin{figure}[h!]
    \centering
    \includegraphics[width=0.95\linewidth]{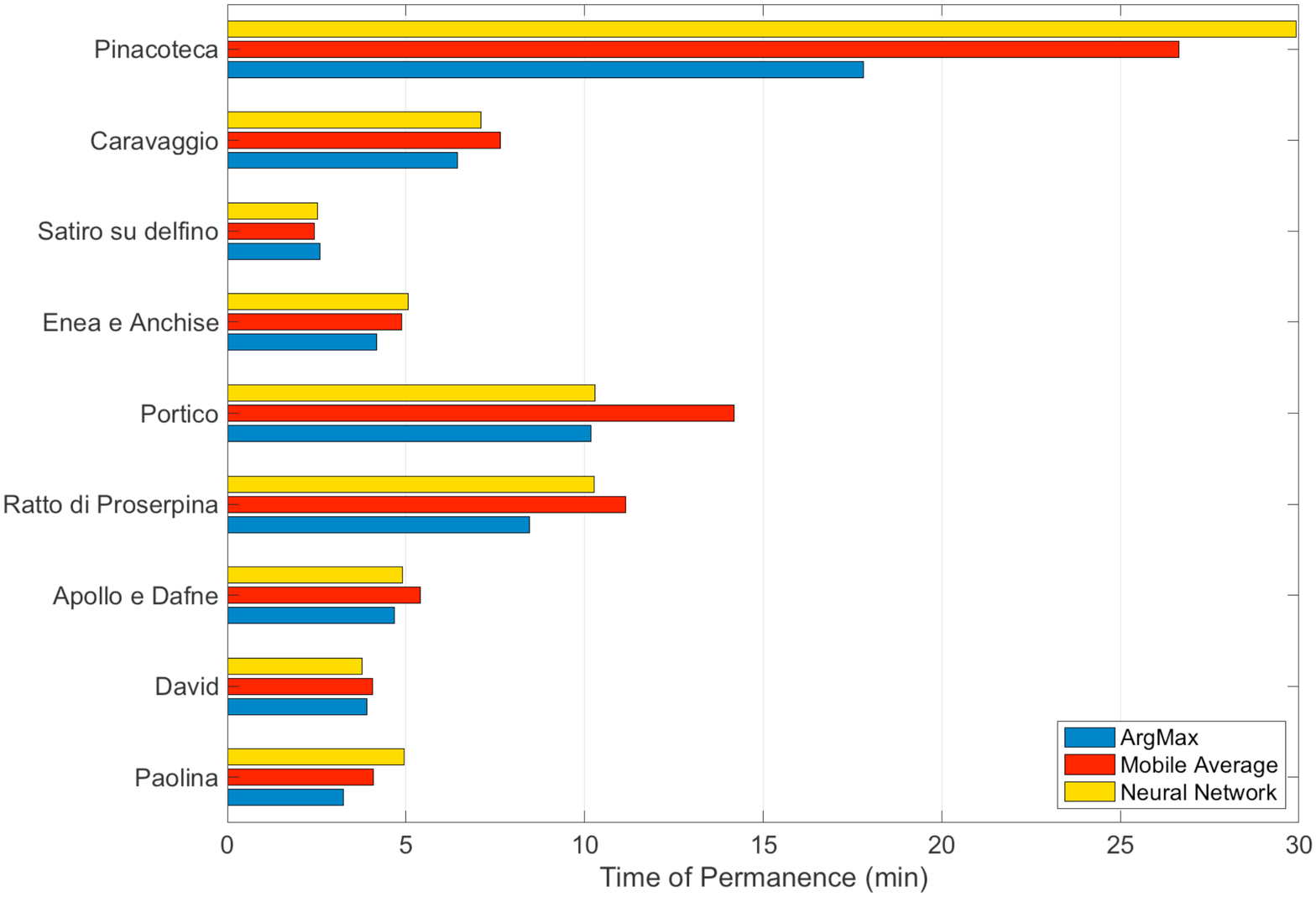}
    \caption{
        Room-scale Time of Permanence evaluated using AM, MA and NN approaches.
    }
    \label{F:ToP-comparison}
\end{figure}

Besides this accuracy measure, looking at single trajectories we observed that the MA approach is more efficient in catching the transition between rooms, while NN is more precise in evaluating the Time of Permanence inside each room, see Figure \ref{F:ToP-comparison}.

\begin{figure}[h!]
    \centering
    \includegraphics[width=0.95\linewidth]{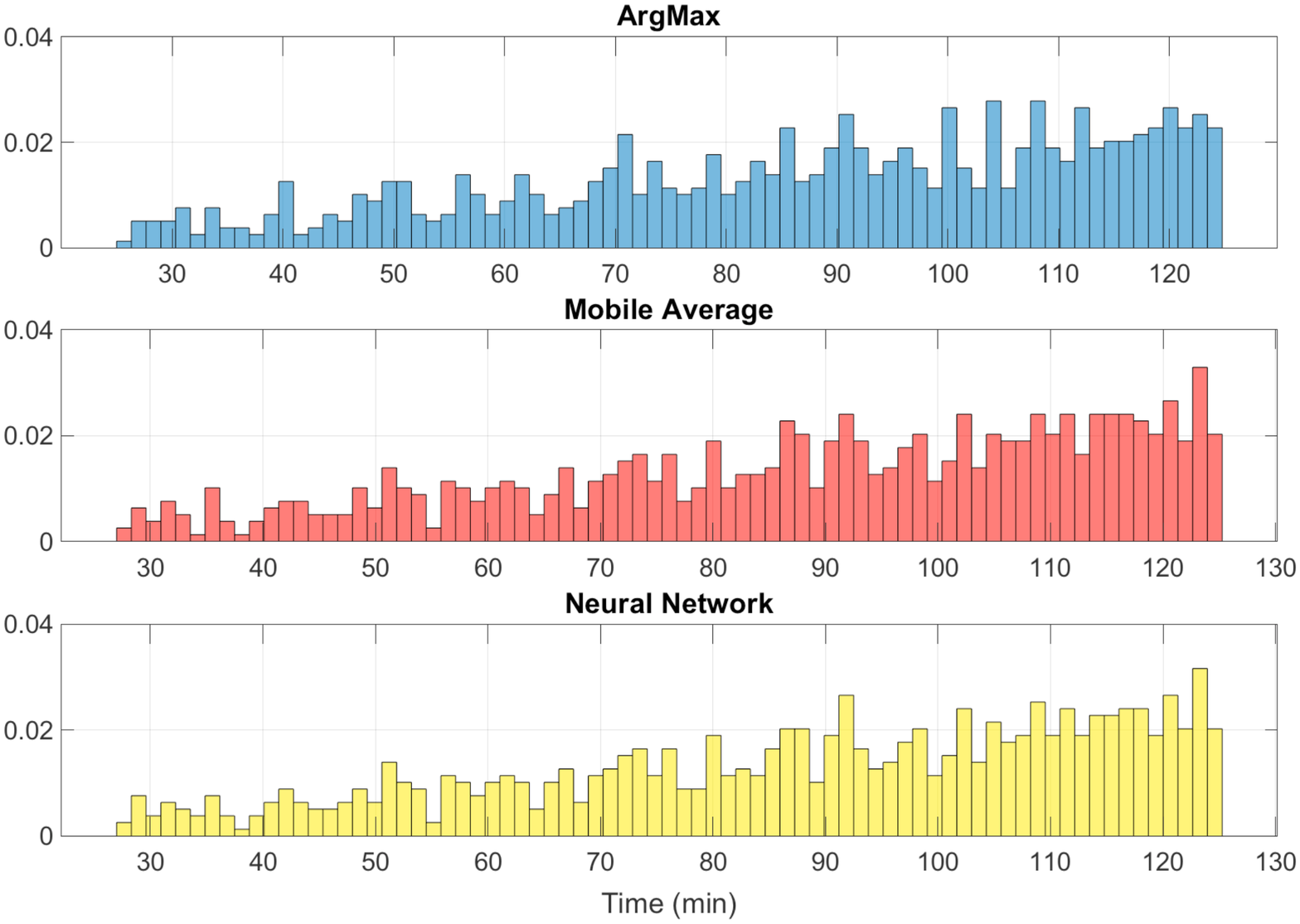}
    \caption{
        Frequency distribution of Time of Visit, for AM, MA and NN approaches. AM evaluates slightly shorter visits due to the lack of a complete RPi covering in \emph{Pinacoteca}. The other methods auto-fill the missing information. On the contrary, MA Time of Visit is a little longer since entrance and exit are rounded more softly, causing an overflow of information.
    }
    \label{F:ToV-comparison}
\end{figure}

Lastly, Figure \ref{F:ToV-comparison} and Table \ref{T:ToV} offer a comparison between the total Time of Visit measured via the three methods we have presented. 

\begin{center}\begin{table}[h!]
    \caption{Average Time of Visit with corresponding standard deviation evaluated in minutes. Trajectories have been gathered together according to visiting experience.}
    \begin{tabular}{c c c c} 
        {\small Visit type / method} &   {\small AM} &   {\small MA}  &   {\small NN}  \\ 
        \hline
        {\small Normal}     &   $86\pm26$       &   $88\pm26$               &   $88\pm26$               \\
        {\small Audioguide}      &   $102\pm17$      &   $103\pm16$              &   $103\pm16$              \\
        {\small Guide}      &   $103\pm12$      &   $105\pm12$              &   $105\pm12$              \\
        \hline
    \end{tabular}
    \label{T:ToV}
\end{table}\end{center}%
\section{Trajectory analysis}
 Our sample set consists in 900 trajectories: 57 for audio-guide, 24 for guide and 819 for normal visit. We excluded from our initial dataset visits which last less than 25 minutes and more than 125 minutes, this anomalies being probably caused by beacons malfunctioning or not returned. 
 Once a consistent set of suitable trajectories is available, several statistical tools can be applied to obtain information about the visitors' flow.
For example, trajectories can be used to understand the spontaneous tendencies of the crowd flowing inside the museum. With an accurate screening it is possible to figure out if some area of the museum is less visited or even less appreciated by visitors than others. This could point out where improvements may have a higher impact on the visitors' experience.

We now present three quantities of interest we have determined as key points in order to understand how people behave in the museum.
Note that the second one, the \emph{Clockwisety}, is specifically conceived for the Galleria Borghese, due to its particular structure, while the others are general tools which could be used in any environment.

\subsection{Time of Permanence} 
Different rooms have likely different permanence time. This is an expected trend since visitors spent more time in rooms which houses famous paintings. This factor can lead to overcrowding problems which not only make the experience less pleasant for visitors but also, and more importantly, expose masterpieces to risky conditions such those caused by human transpiration and respiration.
Our analysis gives a useful tool to identify this kind of rooms, to appraise the mean overcrowding, and, consequently, setting the best working conditions of the air conditioning system, the entry rate of visitors, and the pieces of art positions.

\begin{figure}[t]
    \centering
    \includegraphics[width=0.95\linewidth]{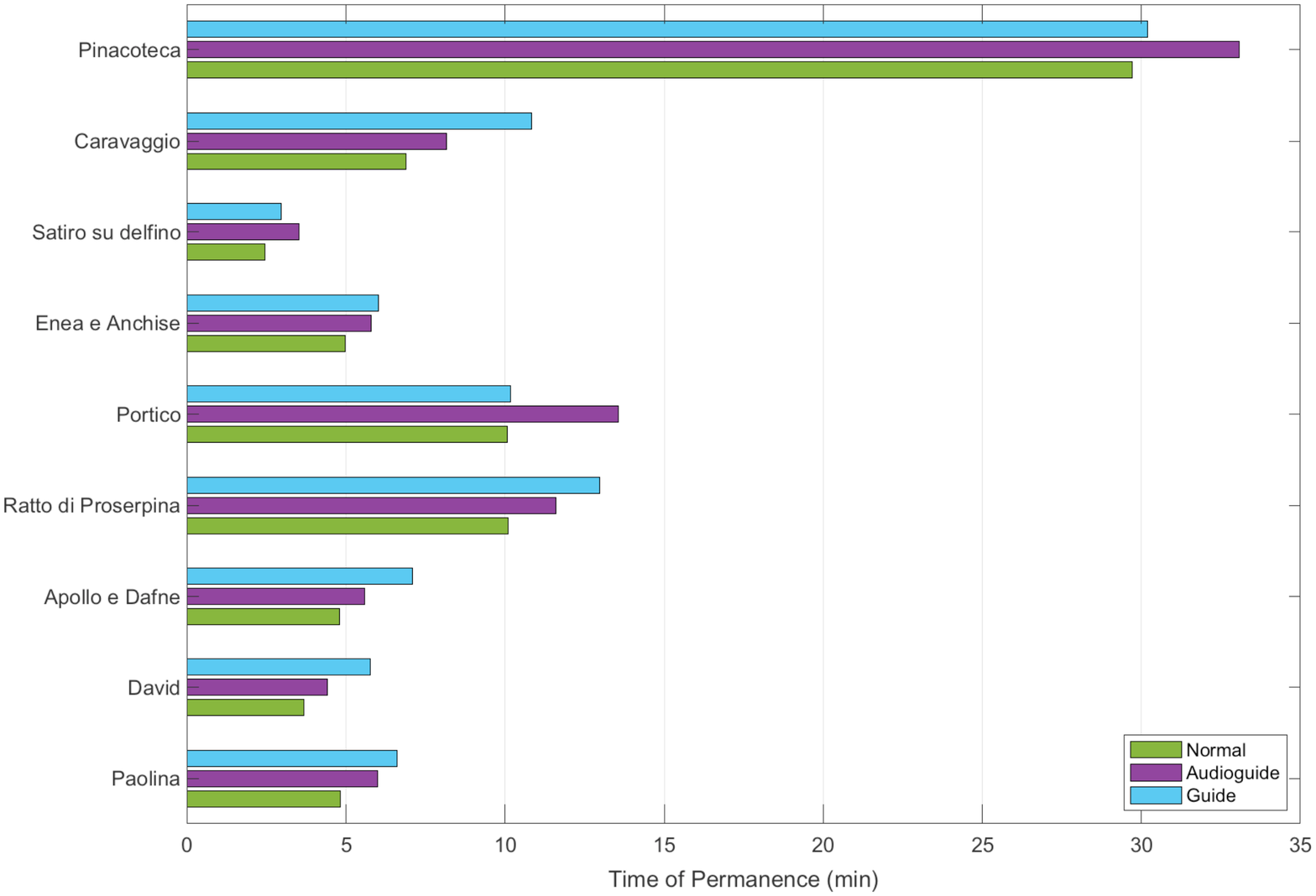}
    \caption{
        Average room-scale Time of Permanence sorted according to the visiting experience (NN method). 
    }
    \label{F:ToP-NN}
\end{figure}

From Figure \ref{F:ToP-NN} we can see that Guided tours spend more time in rooms hosting renowned masterpieces like Caravaggio's paintings and Bernini's \emph{Ratto di Proserpina}; those who are audio-guide equipped, spend, on average, more time in Pinacoteca and Portico, since many more artworks are included in the audio tour from this rooms.

In addition, from Table \ref{T:ToV}, we can see that guided tours present longer time of visit as well as a smaller variance.
This trend is coherent with the fact that usually tours follow similar patterns and have approximately the same length.

\subsection{Clockwisety}
The floor plan of Galleria Borghese suggests us a quantity to measure:
since the first floor of the museum can be considered as a closed cycle with two antipodal access points (the \emph{Portico} and \emph{Ratto di Proserpina}'s room), we measure the ``clockwisety'' of a trajectory. We assigned to each trajectory a score based on how many doors it goes through counterclockwise ($+1$) and how many are crossed clockwise ($-1$).
For example, a visitor who performs a complete counterclockwise tour from the \emph{Portico} and back, and then goes clockwise to the \emph{Pinacoteca}. Since s/he goes through $8$ doors counterclockwise and $4$ doors clockwise, his/her clockwisety will be equal to $4$.
Note that $\pm 8$ is the score of a single complete tour of the first floor.

\begin{figure}[t]
    \centering
    \includegraphics[width=0.95\linewidth]{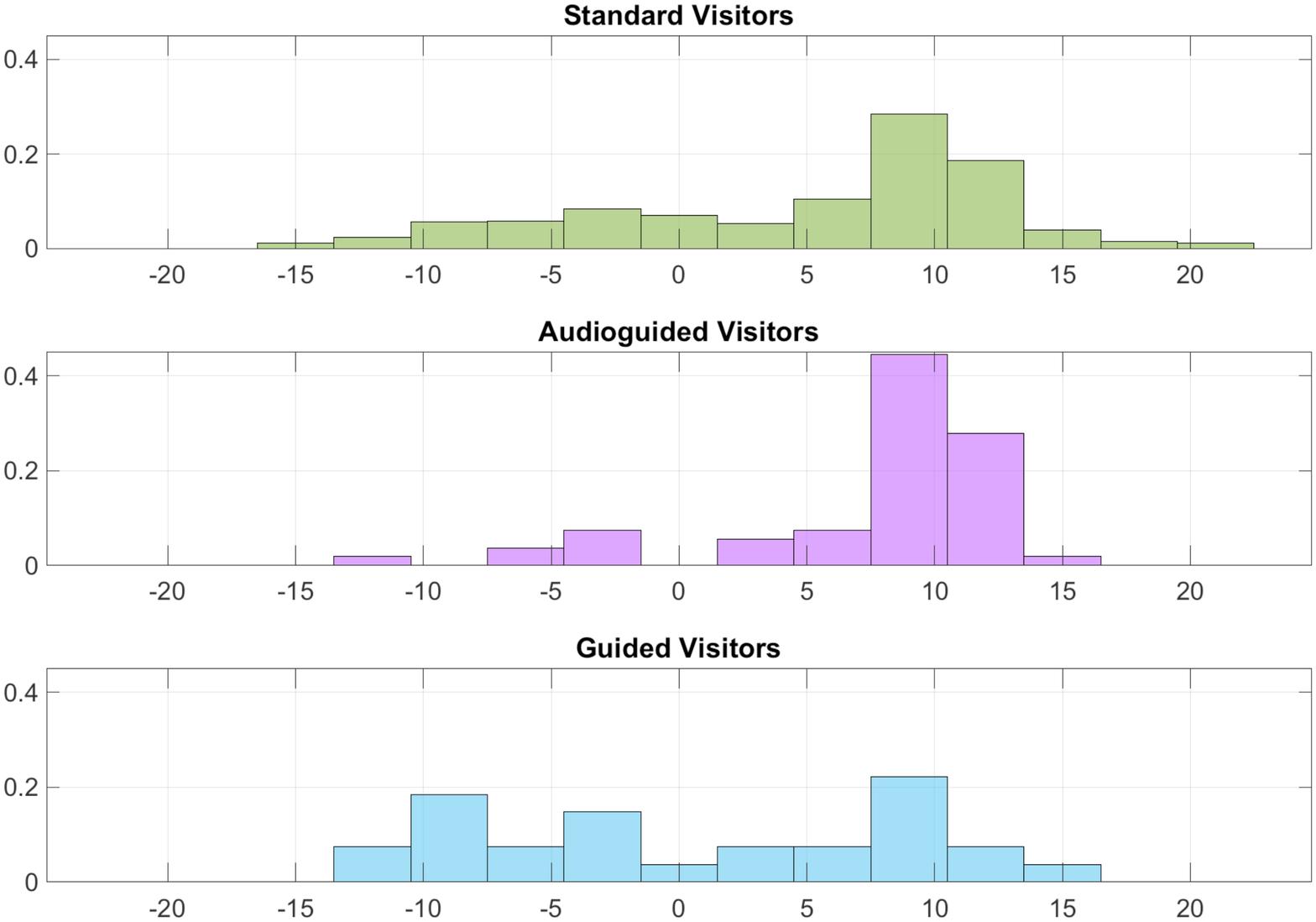}
    \caption{
        Percentage of the clockwisety distribution for the three visiting experiences. 
    }
    \label{F:CW}
\end{figure}

Figure \ref{F:CW} shows the clockwisety distribution for all trajectories for the three visitors experiences.
We can observe that Audioguide trajectories, more than others, are efficient visits with a single counterclockwise round.
On the other hand, a major degree of uniformity on the clockwisety distribution amongst the guide trajectories can be legitimated by a search of less crowded visit paths.
\subsection{Categorization and Clustering}
Beside the elementary tools we have described, more sophisticated tools can be applied in order to obtain trajectories' categorization and clustering.
To this aim, the main ingredient we need is a suitable notion of distance between trajectories, that is a \emph{metric} which allows us to quantify how much close or far two trajectories are from each other.

In order to define it, we were inspired by an advanced and powerful tool from the optimal transport theory, namely the Wasserstein distance: often referred as earth mover's distance, it is usually used to quantify the distance between two density functions.
Roughly speaking, in our case the idea is to quantify how much it costs to transform one trajectory -- bin by bin -- to the other, until they overlap.

The building block is the definition of a cost function to weight the distances between each couple of rooms $(r_i, r_j)$ via a function $w(r_i, r_j) \to \mathbb C^+$, with $w(r_i, r_j) = 0 \iff i = j$.
To build such a weight function we fixed, for the rooms inside the museum, a weight of $1$ between connected rooms and then we added a weight of $2$ for each room a visitor has to pass through to reach the other room following the shortest path inside the museum; the only connections that break this rule are the ones with the \emph{Pinacoteca}, which are weighted $15$ plus the cost for reaching the stairs (in \emph{Ratto di Proserpina}). For the connection with the outside of the museum we fixed a weight of $10 i$. 
The imaginary part is used in order to preserve the difference between two trajectories that are both inside the museum or not; the factor 10 beside the imaginary part, is chosen in order to balance the two values if combined with the euclidean norm, as we did in the rest of this paper.

\begin{figure}[t!]
    \centering
    \includegraphics[width=0.95\linewidth]{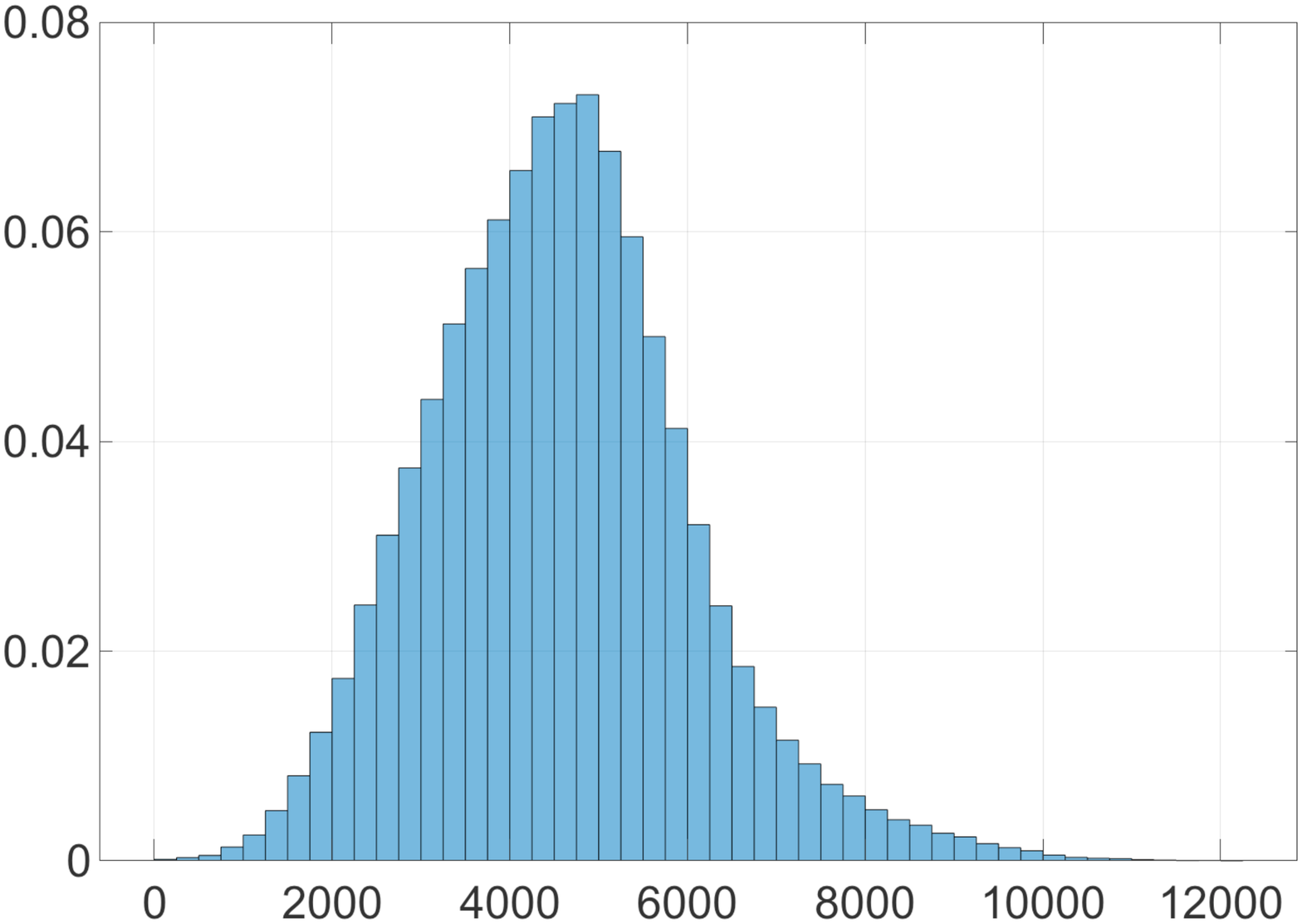}
    \caption{
        Frequency distribution of all-to-all mutual distances between trajectories in the dataset.
    }
    \label{F:plot900x900}
\end{figure}

The last step is to integrate in time the distance between the rooms occupied by the two visitors, thus obtaining the metric $d$ defined as
\begin{equation}\label{def:d}
    d(X,Y) := \sum_{t=1}^m w(X^{(t)}, Y^{(t)}),
\end{equation}
where $X$ and $Y$ are the two trajectories (\emph{i.e.} sequences of rooms), $t$ spans over the time bins, and $X^{(t)}$, $Y^{(t)}$ are the rooms occupied by the two visitors at time bin $t$, respectively.
This metric can be used in many ways, combining the real and imaginary part with different weights in order to catch different features.

\begin{figure}[t!]
    \centering
    \includegraphics[width=0.95\linewidth]    {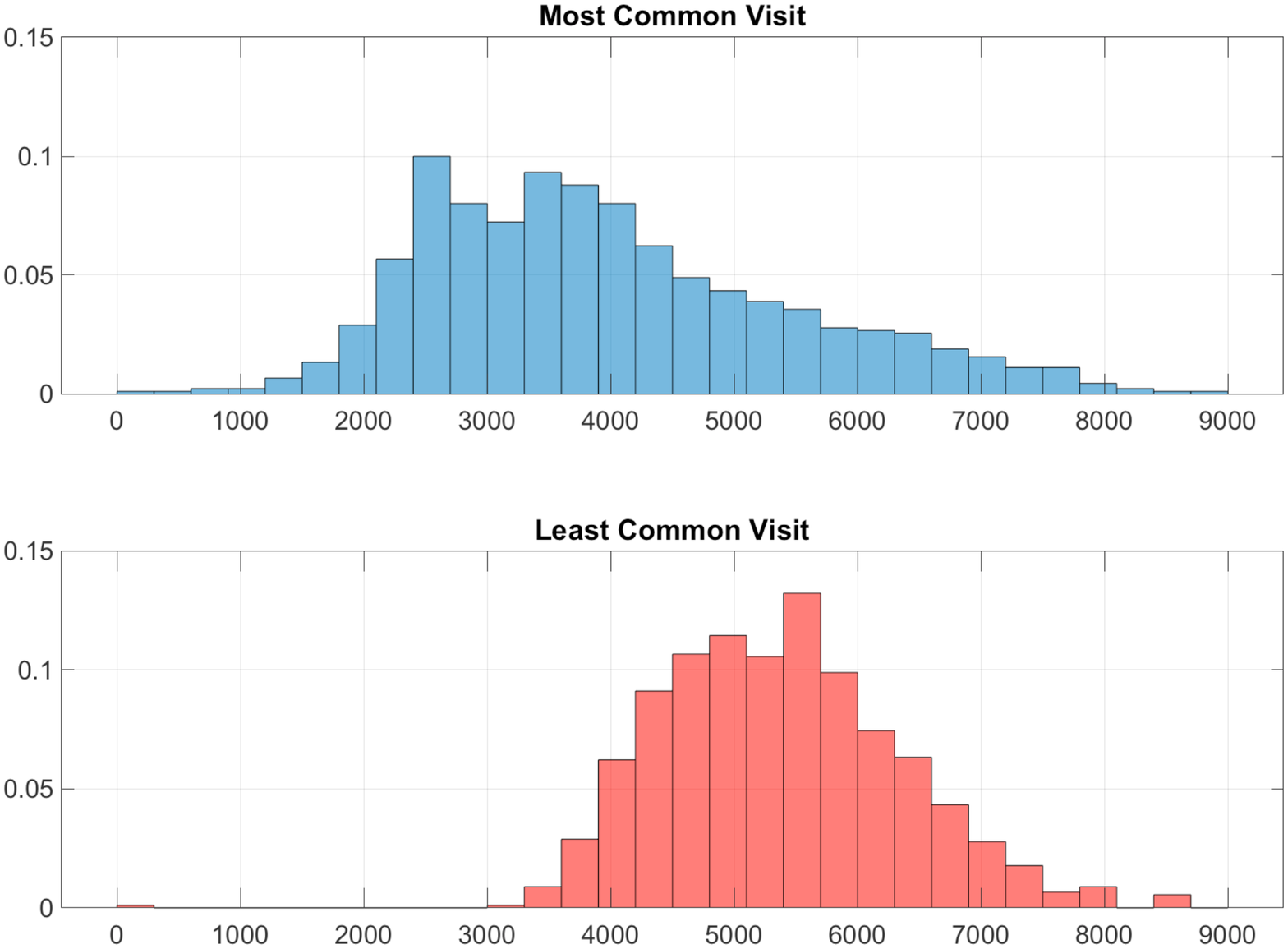}
    \caption{
        Frequency distribution of the distances $d$ from the most (top) and least (bottom) common visit path to all the others trajectories. Corresponding trajectories are displayed in Figure \ref{F:common-uncommon-traj}.
    }
    \label{F:common-uncommon-graph}
\end{figure}

\begin{figure}[t!]
    \centering
    \includegraphics[width=0.95\linewidth]{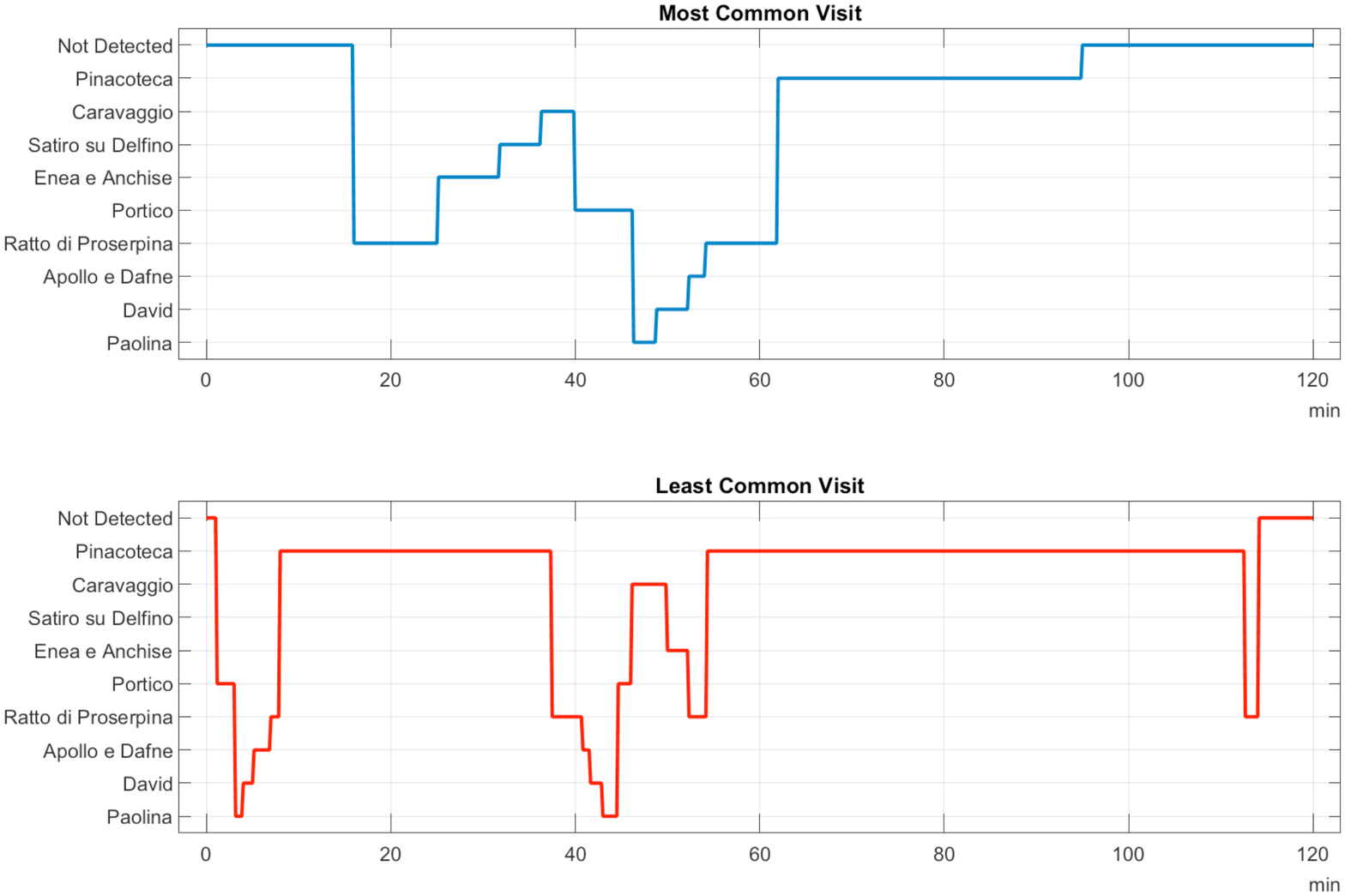}
    \caption{
        Trajectories identified in Figure \ref{F:common-uncommon-graph}.
        The most common visit (top) starts from \emph{Ratto di Proserpina}'s room and goes to the \emph{Pinacoteca} as last step before leaving the museum. It has, as expected, the clockwisety value equal to $8$. 
        The least common visit (bottom), instead, starts from the \emph{Portico}, rushes counterclockwise to the \emph{Pinacoteca}, gets back to the first floor (where follows a clockwise fast visit) and gets back (for the second time!) to \emph{Pinacoteca}. Practical experience confirms that it hardly happens that any trajectory goes twice up and down the stairs, corroborating this result.
    }
    \label{F:common-uncommon-traj}
\end{figure}

Metric $d$ can be used to find the most and least common visit path. To do so, we first compute all the mutual distances between all the trajectories in the dataset, getting a $900\times 900$ symmetric matrix. The histogram of its entries is shown in Figure \ref{F:plot900x900}.
Evaluating the most common interval (\emph{i.e.} the interval mode) of distances from each trajectory to each other and taking the infimum (resp., supremum) leads to the most (resp. least) common trajectory. The results of this computation are shown in Figures \ref{F:common-uncommon-graph}-\ref{F:common-uncommon-traj}.

\begin{figure}[t!]
    \centering
    \includegraphics[width=0.95\linewidth]{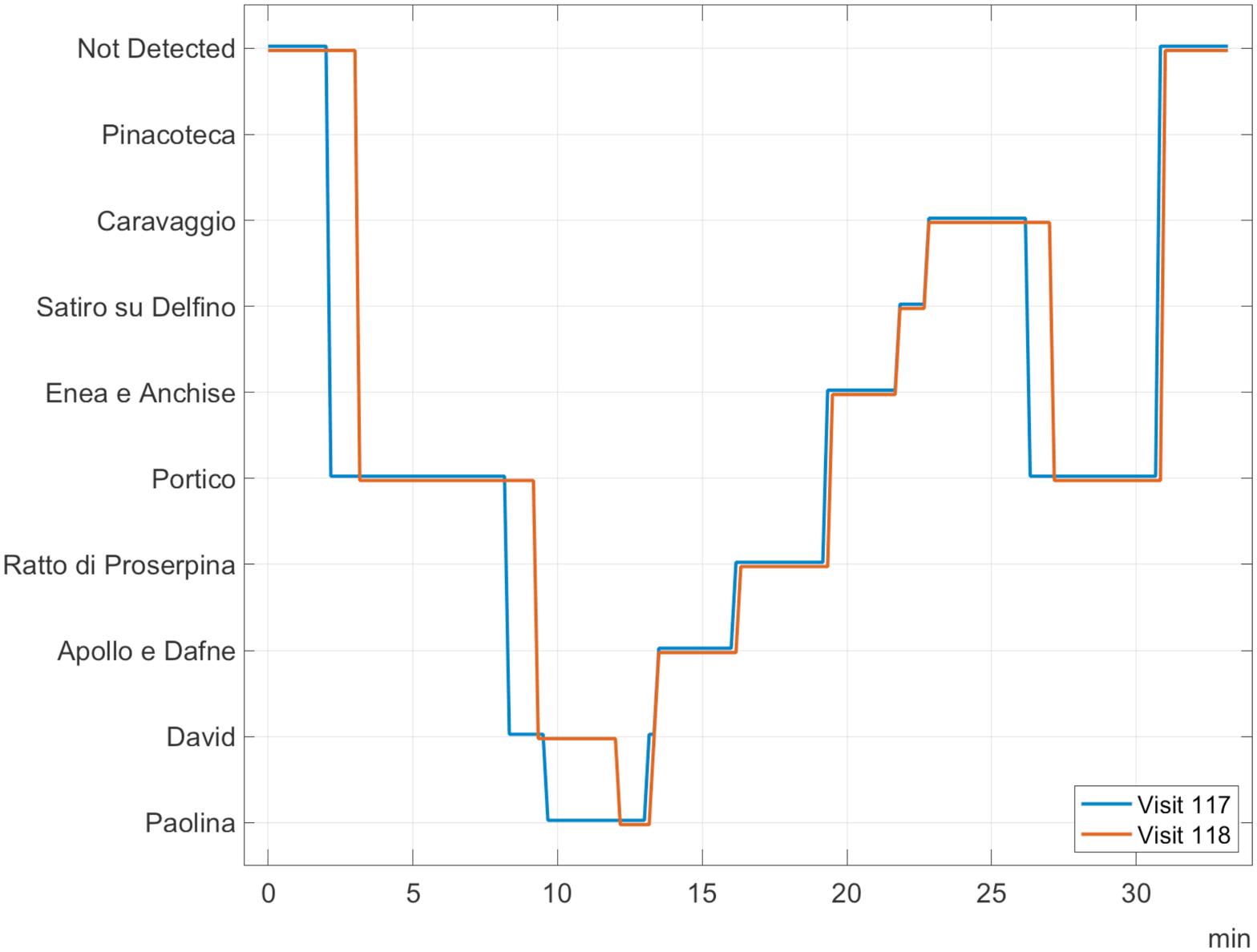}
    \caption{
        Trajectory of two members of the same family visiting together the museum with a distance, according to \eqref{def:d} of $90.3$.
        As it could be seen, beacon $117$ is leading the visit as the transitions of its trajectory occur earlier than those of the beacon 118.
    }
    \label{F:TC}
\end{figure}

\begin{figure}[t!]
    \centering
    \includegraphics[width=0.95\linewidth]{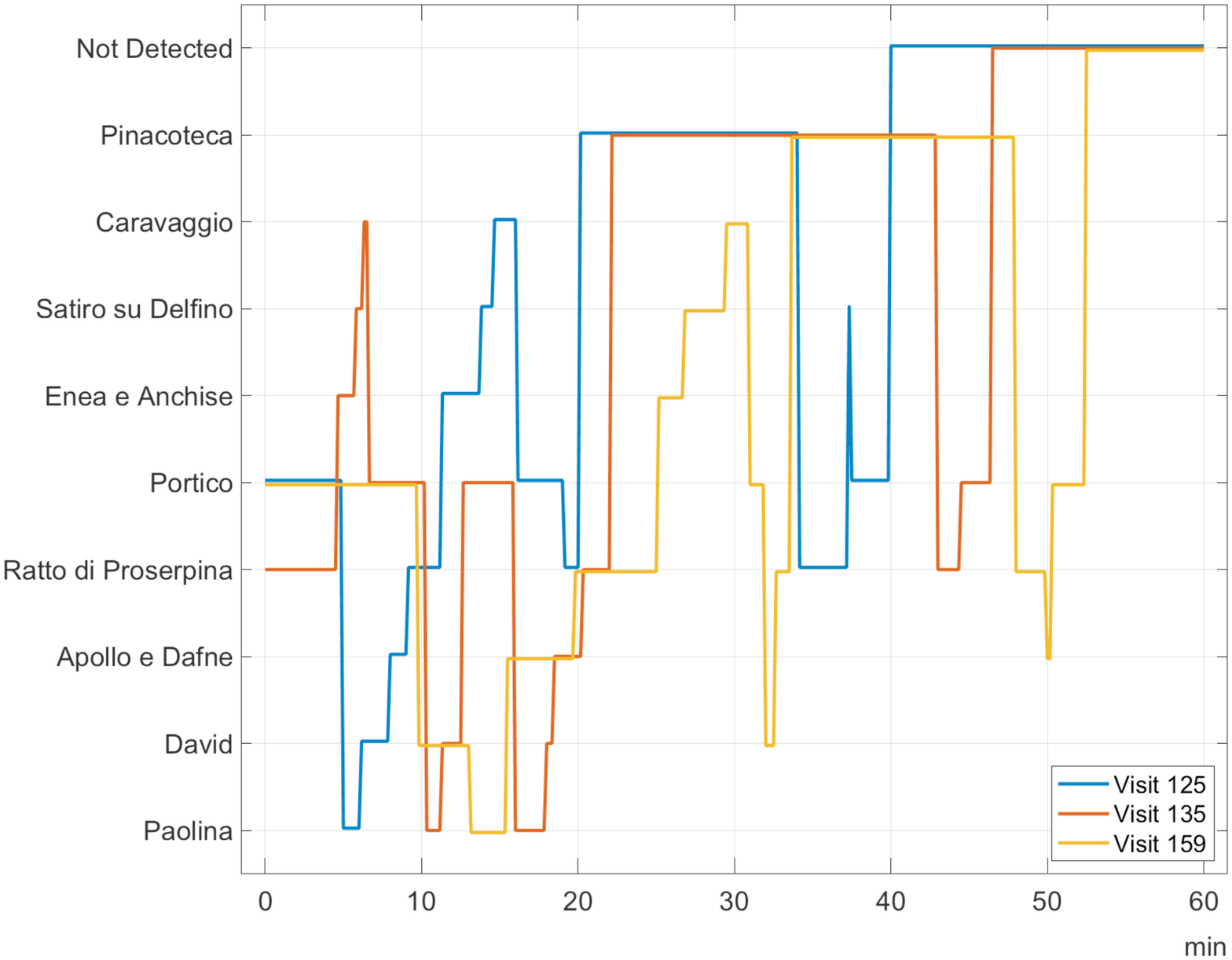}
    \caption{
        Three independent visitors trajectories. Even if all the trajectories follow approximately the same path, their mutual distance is higher than $1500$. Due to the high similarity during \emph{Pinacoteca}'s visit, trajectories $125$ and $135$ are relatively close (distance of $1614$). On the other hand, trajectory $159$ achieves a distance of $2015$ from trajectory $135$ and a distance of $2568$ from trajectory $125$. 
    }\label{F:TD}
\end{figure}

Moreover, the mutual distances can be also used to quickly identify the trajectories belonging to people moving together (in group), even without knowing \emph{a priori} that such a persons belong to a group. In fact the metric $d$ defined in (\ref{def:d}) gives very low values when trajectories of people moving in the same group are evaluated. An example is reported in Figure \ref{F:TC}, where two trajectories have been detected to be of two members of the same family (with mutual distance of $90.3$).
Figure \ref{F:TD} shows, instead, three independent trajectories whose mutual distance is higher than $1500$.

This statistical analysis can be also used for the automatic detection of guided tours extraneous to the organization of the museum.
    
\section{Future work}
The research tools introduced above can be used for many future goals.
The analysis of the trajectories can be used to evaluate 
the transition matrix which define the probability $P^X_{i,j}$ to switch from room $i$ to room $j$ assuming a visit $X$ has occurred. The knowledge of this matrix is fundamental to understand how people behave in the museum and, from a statistical point of view, allows us to define a Markov chain with memory through which we can model the crowd flowing and the visitors' movements. 

Once the transition probability from one room to another has been extracted (also taking into account the history of the visit, \emph{i.e.} which rooms are already visited and for how long), it is possible to build a previsional model, a sort of \emph{digital twin of the museum}, to simulate the occupancy rate of the rooms. More precisely, giving in input the number of visitors, their entry time and the entrance employed (if more than one), the simulator can forecast the trajectory of each visitor, also considering the impact of the interaction between all of them (\emph{e.g.}, what happens if a visitor desires to enter a room very or totally congested), cfr.\ \cite{Ratti2}. 
A simulator can therefore be developed in order to study the impact of changes to the museum management, like, \emph{e.g.}, suggested visit path, entry conditions, maximum number of visitors allowed, time limits, closure of some rooms and many others. 

Last but not least, visitors' trajectories can be put in relationship with environmental parameters such as CO$_2$, temperature and humidity. This allows to assist intelligent air conditioning system to preserve desired values in each room, taking into account the presence of visitors.

\end{document}